# Naming is Framing: How Cybersecurity's Language Problems are Repeating in AI Governance

Lianne Potter

## Abstract

Language is not neutral; it frames understanding, structures power, and shapes governance. This paper argues that misnomers like cybersecurity and artificial intelligence (AI) are more than semantic quirks; they carry significant governance risks by obscuring human agency, inflating expectations, and distorting accountability. Drawing on lessons from cybersecurity's linguistic pitfalls, such as the 'weakest link' narrative, this paper highlights how AI discourse is falling into similar traps with metaphors like 'alignment,' 'black box,' and 'hallucination.' These terms embed adversarial, mystifying, or overly technical assumptions into governance structures. In response, the paper advocates for a language-first approach to AI governance: one that interrogates dominant metaphors, foregrounds human roles, and co-develops a lexicon that is precise, inclusive, and reflexive. This paper contends that linguistic reform is not peripheral to governance but central to the construction of transparent, equitable, and anticipatory regulatory frameworks.

## 1. Introduction

Language is not just a tool of communication, it is a tool of power. This paper explores how language, especially how we name technological disciplines, shapes how we understand them and affects how we govern emerging technologies. Specifically, it examines how the terms 'cybersecurity' and 'artificial intelligence' (AI) have led to fundamental misunderstandings that undermine effective governance. These names do not just mislead. They obscure human responsibility, inflate or distort expectations, and fracture interdisciplinary collaboration. Drawing lessons (and missteps) from cybersecurity, this paper argues that misnomers are more than just semantic slights; they are governance risks. If we hope to build regulatory frameworks that prioritise safety, transparency, and accountability, we must start by choosing our words more carefully. Naming is not just a description. It is about power. Therefore, reframing the language may be the first step to reframing the future.

This paper proceeds in four stages. First, it examines the governance failures embedded in cybersecurity's linguistic framing, focusing on how terms like 'weakest link' contributed to adversarial dynamics and misplaced blame. Second, it maps emerging linguistic risks in AI discourse, including metaphors such as 'alignment,' 'black box,' and 'hallucination,' and explores their governance implications. Third, it proposes a language-first approach to AI governance; one that critically interrogates metaphor, recentres human agency, and reframes AI as a socio-technical system. Finally, it advocates for the co-creation of inclusive governance vocabularies through participatory processes, highlighting the importance of clarity, accessibility, and cultural sensitivity in building anticipatory and democratic governance structures. In doing so, the paper contends that linguistic framing is not a peripheral concern but foundational to how we regulate and relate to technological systems. As we hurtle toward an AI-governed future, we risk repeating the same linguistic and conceptual mistakes we made with cybersecurity, only this time, with far greater consequences, so now is a perfect opportunity to learn from our sister discipline.

## 2. "'Tis but thy name that is my enemy"

What is in a name? Juliet famously asks this question in Romeo and Juliet, pleading to discard the labels that keep her from love. But names, as it turns out, are powerful. They shape how we understand the world, relate to each other, and construct systems of power and control. "A rose by any other name would smell as sweet," (Shakespeare, 1899, p. 34) Juliet says, but had Romeo been named something else, perhaps the tragedy might have been avoided.

Names carry assumptions, histories, and reputations. They can make or break trust, shape policy, and define entire fields of knowledge. Once named, a thing is framed—often for good. Take the name 'Karen,' for example. A once innocuous name meaning 'pure' (Greek), it now conjures a meme-driven stereotype: that of a self-entitled, middle-aged woman, demanding to speak to the manager. While this shift may seem trivial or humorous, its cultural impact is measurable. In the US, the name 'Karen' plummeted in popularity following the meme's rise in 2019, dropping by 162 points in 2020 and a further 427 in 2021 (see Figure 1). Parents could not have predicted this cultural shift, but it illustrates a deeper truth: names evolve, and with them, perceptions.

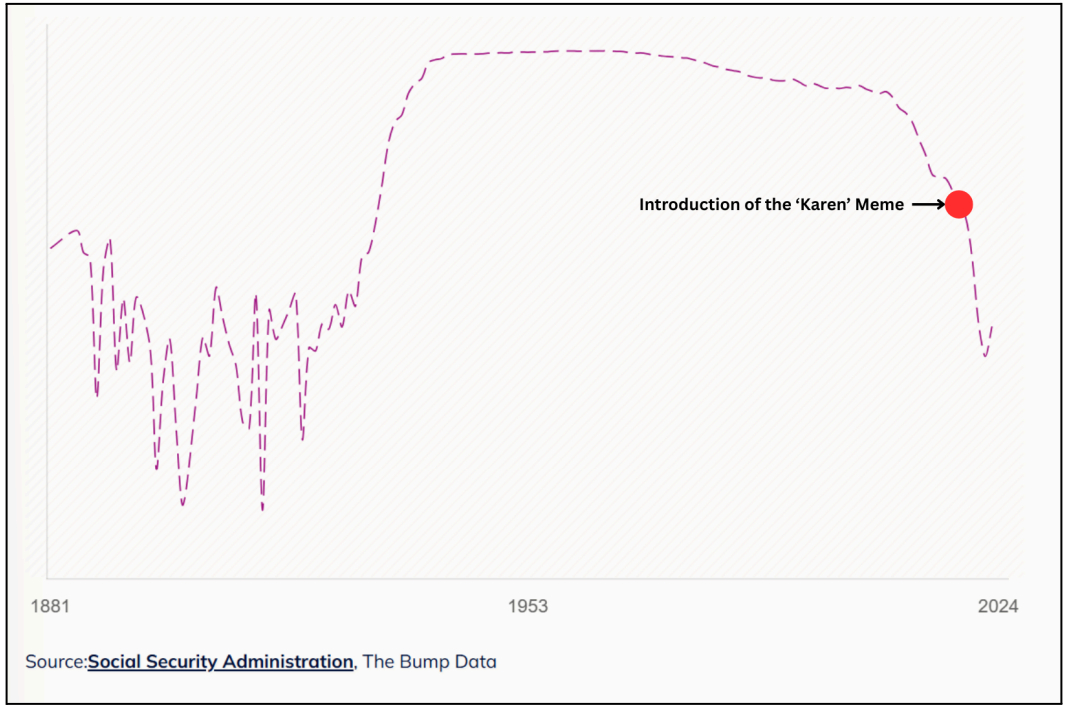

*Figure 1: Popularity of the Name 'Karen' by Year in the US and the Year the 'Karen' Meme Appeared in Google Searches, Author's Addition (Whelehan, 2025)*

Just as Karen now triggers an image far removed from its original meaning, so too does the name cybersecurity carry emotional baggage. For some, it invokes fear of hackers and breaches, or systems failing. For others, it signals obstruction: the team that says "no" to innovation. The name, cybersecurity, shapes how the discipline is understood and engaged. This is no minor branding issue. As we will explore, language like this does not just misrepresent the field, it undermines collaboration, warps governance, and fuels misunderstanding - and AI may be heading down the same path.

## 3. Misnomers Are Not Just a Semantic Problem; They Are a Governance Risk

One of the biggest misnomers in technology discourse is the term "cybersecurity." Cybersecurity is usually defined as the practice of protecting digital assets (networks, devices, data, software, etc.) from unauthorised, unintended, or criminal access (Stevenson, 2015). But framed that way, it sounds like an activity that happens "out there" in the cloud, in code, or the nebulous space known as 'the Internet.' Security is not, however, a destination but a best endeavour.

The term 'cyber' implies that cybersecurity is only concerned with technology, removing the human from the loop. However, cybercrime is profoundly human. Its victims, attackers, and, most critically, the weaknesses that allow it to succeed, are human. This techno-centric view obfuscates cybersecurity's fundamentally human nature.

If the word 'cyber' misdirects us from the human, then 'security' gives us false reassurance. While 'cyber' promotes detachment, 'security' offers an illusion of safety that misrepresents the perpetual nature of digital threats. 'Security' implies safety, stability, and a protected status that suggests that once achieved, it needs only to be maintained. But anyone paying attention to mainstream news knows this is not the case, and that even the most sophisticated technology companies suffer regular breaches (Edwards et al., 2016; Makridis, 2021).

Most security professionals today operate on the premise of 'assumed breach.' That is, they act as if an attacker is already inside the system, just undetected. Instead of a binary 'secure or not,' modern cybersecurity is about constant risk management, detection, and response. This shift has reshaped how cybersecurity teams work. Working in cybersecurity requires constant vigilance, but working in this way also breeds tension, stress and paranoia, which can affect decision-making and organisational culture (Mizrak et al., 2025; Rodriguez-Bermejo et al., 2021; Singh et al., 2023).

The language we use to define our speciality matters because how people perceive us has far-reaching impacts on how our discipline is implemented, especially from a governance perspective. Governance requires a cross-functional understanding, which is difficult to achieve if there are misconceptions about its potential. The words we use shape the policies we create. Misnomers are not just a semantic problem; they are a governance risk. Moreover, if something is misnamed, it can confuse people, and people can assume it is someone else's problem.

Framing cybersecurity as purely technical misunderstands its nature. It is as much about anthropology, behavioural science, psychology, economics, and philosophy as it is about software and hardware (Creese et al., 2021; Jeong et al., 2021). By understanding that cybersecurity is as much about human collaboration and culture as it is about technology, we gain a valuable lens for developing AI governance that prioritises humanity at its core.

## 4. What did you Just Call Me?! The Weakest Link

In Britain during the early 2000s, a single phrase could be heard parroted in every pub and playground throughout the country: 'You are the weakest link. Goodbye!' Delivered with icy precision by Anne Robinson on the game show *The Weakest Link*, the line became a cultural shorthand for failure. No one wanted to be the weakest link, not in a gameshow, and certainly not in real life either. Yet decades later, the phrase remains a mainstay of our cybersecurity vernacular. Users, especially those outside the security profession, are still routinely described as the 'weakest link' in the chain. It is a sentiment born of frustration, but it is also deeply counterproductive. To those on the receiving end, it is more than a quip, it is an indictment. It suggests that users are the root of all breaches rather than co-participants in a shared security endeavour. This is not the only phrase that alienates in cybersecurity: 'insider threat,' 'least privilege,' 'zero trust,' these terms frame people as risks to be mitigated rather than partners to be empowered (Sasse et al., 2001).

This dynamic can create a culture of fear and blame rather than one of curiosity, support, and empowerment. If we want people to care about cybersecurity, we have to meet them where they are. That means using language that invites participation, acknowledges shared responsibility, and recognises that human behaviour is complex, not simply a flaw to be patched or a vector to be monitored. Cybersecurity is not just a technical challenge; it is a social one. And if we are serious about creating resilient systems, we need to stop asking, 'Who's the weakest link?' and start asking: 'How do we make every link stronger?' After all, a chain is only as strong as the relationships between its links and those are forged with empathy, not blame.

The parallels between cybersecurity and AI become even more striking when we consider the role of language. Just as cybersecurity has long suffered from combative, alienating terminology, so too does AI risk falling into a similar trap. Terms like 'alignment,' 'control problem,' and even 'black box' frame the human-AI relationship in adversarial terms (Almada, 2023; von Eschenbach, 2021). They suggest that AI is something dangerous and unpredictable that needs to be controlled, monitored, or contained, much like how users in cybersecurity are viewed as unpredictable variables to be locked down and limited.

This kind of language subtly encodes power dynamics into technical systems. It paints a picture where trust is absent, where the human is either a threat to the system or at risk from the system, and where relationships are transactional, not collaborative. If we want to build responsible, trustworthy AI systems, just like we want secure digital environments, we need to be conscious of how language shapes design. Words are architecture. They tell people whether a system is for them or against them. Whether they are welcome participants or merely tolerated sources of noise. Perhaps it is time for both cybersecurity and AI to update their lexicons. To move away from frameworks of control, blame, and exclusion, and toward language that reflects shared stewardship, mutual understanding, and human-centred values. If AI is going to reflect our world, we should be careful about what we are saying.

## 5. What's So 'Artificial' About Intelligence?

Like cybersecurity, the term 'artificial intelligence' creates misconceptions that hinder its ability to deliver effective governance and safety. Due to its representation in popular culture, the term can sometimes conjure the image of a machine that can think, learn, and reason like a human and maybe even one day achieve sentience. This framing can lead to a misunderstanding of what AI *actually* is: an intricate system of algorithms, data processing, and learning models rather than something akin to human intelligence.

The challenge occurs, particularly as we develop governance frameworks, when we put too much stock into AI's 'intelligence' and outsource decision-making to it, trusting it to make the 'right' choices if enough controls have been embedded. But the 'artificial' part of AI is anything but. As with cyber, it is the product of human design and input. AI is a tool created, trained and guided by humans. Not only does it have the potential to reinforce biases because its human creators may reflect their own biases into the algorithm (Fang et al., 2025; Harfouche et al., 2023; Taeigagh, 2025), but it is also dangerous because humans are fallible and may unintentionally (or intentionally) develop vulnerabilities that cause AI safety concerns.

Just as cybersecurity is ultimately about the humans involved (attackers, defenders, victims), AI is human-centric because it involves human choices. Namely, how we define, train, and implement AI, interpret its results, and govern it on a micro and macro scale. Yet, both cybersecurity and AI are often perceived through a technical lens, which leads to misunderstandings, oversights, and, ultimately, human failure.

## 6. The Governance Consequences of Misnaming AI

Understanding that both cybersecurity and AI are deeply human concerns highlights a crucial point: language influences governance. When the terms themselves are misleading, governance becomes reactive rather than anticipatory. If AI is seen as inherently intelligent and neutral, we may be tempted to treat it as a system that will make ethical decisions for us. If it is framed as autonomous, we may fail to assign accountability when things go wrong. If it is perceived as something we can make 'secure,' we may ignore the need for continuous governance. Just as governance in cybersecurity was slow to mature due to a misunderstanding of what was really at stake, AI governance faces the same risk. The technical lens obscures the societal and ethical implications, leading to frameworks that may not be fit for purpose.

## 7. We Are Making the Same Mistakes Again: A Warning from Cybersecurity

Cybersecurity has taught us, arguably the hard way, that no system is ever fully secure, no threat entirely predictable, and no solution ever truly final. Yet when it comes to AI, the stakes are higher, the speed is faster, and the margin for error may be far smaller.

Take the 2017 WannaCry cyberattack, a ransomware cryptoworm that exploited systems that had not installed a recently released Microsoft patch. The attack infected up to 70,000 devices in the UK NHS, including medical equipment. It caused procedures to be cancelled and hospitals to return to manual processes. The cause? Not a lack of care, but systemic issues: legacy equipment, funding limitations, and timing concerns for patching (House of Commons Committee of Public Accounts, 2018; National Audit Office, 2018). We face similar vulnerabilities in AI. People will exploit them, and if we do not learn from cyber, we will not be ready. Because we do not do well at fixing vulnerabilities after the fact, therefore, we need governance that anticipates them.

# 8. A Language-First Approach to Better Governance

We must stop treating 'artificial intelligence' as a mystical force and start recognising it for what it truly is: a human-made tool, embedded with our values, assumptions, and flaws. That reorientation begins with language. The terms we use to describe, categorise, and govern AI are not neutral. They shape how power operates, who is held accountable, and whose voices are centred or sidelined.

A language-first approach to governance recognises that naming is framing (Brugman et al., 2019). It asks policymakers, technologists, and regulators to interrogate the metaphors and terms they rely on, especially those that encode adversarial or hierarchical dynamics. We need governance structures that are transparent and inclusive, and that begins with choosing words that reflect those principles.

I suggest the following as a starting point:

## 8.1 Interrogate the Metaphors

Ask the following question: What metaphors are doing the heavy lifting? Terms like 'alignment,' 'black box,' 'hallucinations,' and 'control problem' conjure up the image of AI as a dangerous, unruly agent. The metaphorical associations of these terms can carry heavy emotional and cultural weight. As a result, they may inadvertently reinforce fear, techno-solutionism, and false binaries (e.g., aligned vs. unaligned AI) or suggest an adversarial relationship. In doing so, they shape how the public understands AI, how policymakers regulate it, and how developers build it.

### 8.1.1 Alignment

For example, the 'alignment' metaphor implies that AI could be something that, with enough calibration, can be nudged into obedience. But alignment presumes a singular, static value set, and often glosses over *whose* values the system should align to, which is why this is often referred to as the 'alignment problem' (Cugurullo, 2024). An alternative to the term 'alignment' might be 'value harmonisation' or 'ethics attunement', suggesting pluralism, negotiation, and continual iteration rather than one-directional control.

### 8.1.2 Trustworthy AI

I would also recommend being wary of the term 'trustworthy AI' (Deroy, 2023). While the goal of creating 'trustworthy AI' is laudable, it is important to recognise that trust is a dynamic and reciprocal relationship, not a fixed attribute. Trustworthiness is often framed as a technical attribute (something that can be engineered or certified). It is, at its core, a relational and context-dependent concept. Trust cannot be mandated; it must be earned, and it is shaped as much by institutional histories and social norms as by system performance. Framing AI as inherently 'trustworthy' risks shifting the burden onto users to accommodate the system, rather than demanding accountability and transparency from developers and deployers, and hence my caution. Instead, governance language should reflect this dynamic by focusing on verifiability, contestability, and agency, terms that centre users' rights to question, understand, and challenge AI systems. By moving away from metaphors that imply obedience or moral character, and towards language that foregrounds interaction, uncertainty, and situated knowledge, we lay the foundation for more democratic and responsive forms of AI governance. Therefore, consider using instead: 'verifiable AI systems', 'accountable and contestable AI', 'reliable and auditable AI', or 'AI supporting user agency', which reflects the need to earn trust through demonstrable properties and user empowerment, rather than implying that trust is an inherent feature of the technology itself.

### 8.1.3 Black Box

Similarly, the term 'black box' conveys opacity and unknowability, implying that AI systems are inherently mysterious. But this metaphor can become a self-fulfilling prophecy. If we tell ourselves AI is a black box, we stop expecting transparency, reducing our demand for interpretable or explainable AI (Duran & Jongsma, 2021). Instead, reframing it as an 'opacity challenge' or a 'transparency deficit' foregrounds design responsibility rather than metaphysical mystery. It signals that explainability is not a philosophical riddle, but an engineering and governance task.

### 8.1.4 Hallucination

Even the now-popular term 'hallucination', used to describe when large language models generate false information, anthropomorphises the model (Barrow, 2024;  Deroy, 2023), implying a kind of dreamlike consciousness or involuntary malfunction. But large language models (LLMS) do not hallucinate; they extrapolate from statistical associations, indifferent to truth or coherence. Some scholars argue that 'bullshit' is a more accurate term—not as vulgarity, but as used in Harry Frankfurt's philosophical sense (Gorrieri, 2024). AI is unconcerned with truth and instead focuses on impression or plausibility. Unlike lying, which requires a relationship to truth, bullshit simply bypasses it. This framing reorients responsibility back to the model's training and incentive structure, rather than mystifying it as a cognitive malfunction. Alternatives like 'statistical fabrication' or 'plausible nonsense' (see Maleki et al., 2024 for more alternatives)  might better capture the nature of the problem without misleading anthropomorphism.

### 8.1.5 Control Problem

Finally, the 'control problem' metaphor positions AI as something to be dominated before it dominates us, reinforcing a worldview of adversarial struggle. Sometimes used interchangeably with the 'AI alignment problem', but although they are closely related, there is a subtle difference in emphasis. Whereas the AI alignment problem is focused on the *inner* workings and goals of AI, the AI control problem takes a broader view, looking at the *outer* relationship and power dynamics between humans and AI. Control is about ensuring the AI hits the target we want and that we can stop it or steer it if it goes off course (whether due to misalignment or other unexpected behaviour). Therefore, instead of 'control problem,' we could adopt terms like 'trust architecture' (not to be confused with cybersecurity's 'zero-trust architecture') or 'governance scaffolding,' which suggest co-creation, mutual responsibility, and system-level thinking. The concept of 'governance scaffolding' can be understood as an extension of 'AI scaffolding,' a term rooted in machine learning and cognitive science that refers to structured, often temporary, support mechanisms designed to assist AI systems in solving complex tasks (Agrawal, 2025).

In governance contexts, scaffolding similarly denotes adaptive, system-level frameworks, such as regulatory instruments, oversight bodies, and participatory mechanisms, that guide the development and deployment of AI technologies. Governance scaffolding is inherently modular, evolving alongside AI systems and societal needs through mechanisms like sandboxes for testing regulations, dynamic compliance frameworks, and collaborative platforms for stakeholder engagement. For example, it might initially mandate data bias audits in healthcare AI while supporting broader ethical transparency frameworks. This approach moves beyond rigid control paradigms and emphasises shared agency, resilience, and trust-building, ensuring governance remains responsive to both technological advancements and societal impacts. Rather than adopting a rigid, top-down control paradigm, governance scaffolding emphasises co-creation, mutual responsibility, and iterative adjustment, aligning with a socio-technical perspective in which AI systems and human institutions co-evolve. This shift in metaphor moves beyond adversarial framings like the "control problem" and instead foregrounds trust, resilience, and shared agency.

Metaphors are not just rhetorical flourishes. They are foundational frames that influence how problems are defined, what solutions are pursued, and who gets to participate in shaping them. If we want governance that is inclusive, responsible, and adaptive, we need to start by asking better metaphorical questions. Let us continue this theme by looking further into how this could be applied to AI governance.

## 8.2 Be Explicit About the Human

As discussed earlier, in both cybersecurity and AI, the role of the human is frequently abstracted, minimised, or altogether erased. We speak of 'automated threat detection' or 'autonomous systems' as though humans are merely bystanders, or worse, obstacles to technological progress. But behind every AI output, every breach, every false positive or model decision, there are layers of human design, deployment, oversight, and omission (Johnson & Verdicchio, 2017). A language-first approach actively resists this erasure by re-inserting the human into every layer of the narrative. It reminds us that it is not 'AI vs. humans' but rather 'humans through and with AI.' This framing transforms AI from a self-willed actor into what it truly is: a socio-technical system shaped by human decisions, incentives, and constraints (Johnson & Verdicchio, 2017, p. 587).

Terminology is key. For instance, rather than describing models as 'autonomous,' we should use 'delegated authority systems,' a term that foregrounds human agency and accountability. Delegation is an act, it implies choice, responsibility, and potential revocability. Likewise, we should be wary of calling models 'intelligent agents' without a clear definition. In high-stakes domains like healthcare, education, or criminal justice, such language risks inflating expectations, confusing roles, and muddying accountability.

A more accurate and transparent descriptor might be 'statistical decision tool.' This makes clear that these systems operate by identifying patterns in historical data, not by reasoning, understanding, or ethical discernment. It demystifies their function and recentres responsibility where it belongs, with the humans who build, deploy, and oversee these systems. This shift is not just rhetorical. It has governance consequences. When we label systems accurately, we frame regulation differently. A 'statistical decision tool' invites questions about training data, statistical validity, and bias mitigation. An 'intelligent agent' invites questions about autonomy, personhood, and agency, which may not even be relevant. To govern AI responsibly, we must speak clearly, define terms explicitly, and refuse to let technical shorthand mask human action. We must stop asking how to control AI and start asking how to govern the humans who shape it.

## 8.3 Co-create the Vocabulary

Contemporary AI governance frameworks frequently rely on terminology that is vague, inconsistent, or excessively technical. Phrases such as 'high-risk' and 'human oversight' in the EU AI Act, while ostensibly reassuring, lack the operational clarity necessary for consistent implementation and enforcement. Similarly, the NIST AI Risk Management Framework (2023) invokes principles such as 'trustworthiness' and 'accountability' without offering substantive guidance on how these values should be interpreted or instantiated across diverse institutional and cultural contexts. This linguistic imprecision creates space for regulatory ambiguity, loopholes, and ultimately, a diminishment of public trust.

However, the challenge is not solely legal or technical, it is fundamentally linguistic. Language structures thought and delimits the range of possible interpretations, responses, and metaphors. When governance vocabulary is opaque, elitist, or exclusionary, it risks rendering well-intentioned policies ineffective and misaligned with the communities they purport to serve. Moreover, such language can inadvertently reproduce existing power asymmetries by privileging those fluent in legal, bureaucratic, or technical dialects, while marginalising other forms of knowledge and expression (Blodgett et al., 2020). To address this, it is imperative to reconceptualise AI governance as a linguistic commons: a shared space where meaning is collectively negotiated rather than unilaterally imposed. This requires terminology that is both precise enough for legal and technical practitioners and sufficiently accessible to the broader public, particularly those communities most impacted by AI systems.

This orientation necessitates co-creation rather than mere consultation (Kinnula et al., 2023; Papagiannidis et al., 2025). Governance processes must deliberately incorporate diverse epistemologies and modes of expression, engaging ethicists, linguists, social scientists, artists, and representatives from historically marginalised communities (Attard-Frost, 2025; Delgado et al., 2023; Ta & Lee, 2023). Participatory approaches, such as naming workshops and deliberative forums, should be embedded in drafting standards and regulations. Moreover, linguistic audits should form part of governance impact assessments, interrogating questions such as: Whose language is being used? Who is excluded? How might meanings shift in different social, cultural, or geopolitical contexts? This is particularly pertinent as AI tools tend to neglect low-resourced languages[1] and dialects (Ta & Lee, 2023).

Cybersecurity provides a cautionary precedent. The early adoption of militarised and exclusionary language ('firewalls', 'threat actors', 'demilitarised zone', 'zero-day exploits', etc.) helped foster a culture of fear, opacity, and gatekeeping. A similar trajectory in AI governance, characterised by terms like 'alignment' or 'control' without sufficient conceptual clarity or public engagement, risks replicating these pathologies.

Language is not a neutral vehicle for policy, it is itself a form of governance. Words do not merely describe systems; they shape their design, function, and social reception (Maas, 2023, p.13). For AI governance to be adaptive, inclusive, and just, it must begin with a deliberate, collective effort to construct a shared and accountable vocabulary.

---

[1] 'Low-resource languages are those that have relatively less data available for training conversational AI systems. In contrast, English, Chinese, Spanish, French, Japanese and more of the European and Western languages are high-resource' (Teo, 2021).

Figure 2: Reframing Common AI Metaphors for Better Governance

| Problematic Term / Metaphor | Metaphorical Implication | Governance Risk | Suggested Alternative | Governance Reframe |
|---|---|---|---|---|
| **Alignment** | AI can be calibrated or nudged into obedience. It assumes a singular, static value set. | Glosses over *whose* values; reinforces false binaries (e.g., aligned vs. unaligned); implies one-way control. | **Value harmonisation, Ethics attunement** | Emphasises pluralism, negotiation, and continual iteration rather than fixed obedience or singular value sets. |
| **Trustworthy AI** | Trustworthiness is a fixed, technical attribute that can be engineered into AI. | Shifts the burden onto users to trust; obscures the relational, dynamic nature of trust; potentially reduces demand for developer/deployer accountability & transparency. | **Verifiable AI systems, Accountable and contestable AI, Reliable and auditable AI, AI supporting user agency** | Emphasises earning trust via demonstrable properties (verifiability, reliability, auditability) and user empowerment (contestability, agency); centres user rights; acknowledges trust is relational, context-dependent and earned. |
| **Black Box** | Frames AI as having inherent opacity, unknowability, and mystery surrounding AI systems. | It can become a self-fulfilling prophecy, reducing demand for transparency and interpretable/explainable AI. | **Opacity challenge, Transparency deficit** | Frames the issue as an engineering and governance task requiring design responsibility, rather than an inherent, unchangeable characteristic. |

| Hallucination | Anthropomorphised AI, suggesting consciousness, intent, or involuntary malfunction. | Mystifies AI outputs; distracts from the model's statistical nature and training data/incentive structures. | **Bullshit (philosophical sense), Statistical fabrication, Plausible nonsense** | Reorients focus on the model's indifference to truth, its reliance on statistical patterns, and the need to address training/incentives. |
|---|---|---|---|---|
| **Control Problem** | Positions AI as an adversary to be dominated or controlled; emphasises power dynamics. | Reinforces an adversarial relationship; potentially limits thinking to top-down control mechanisms. | **Trust architecture, Governance scaffolding** | Promotes co-creation, mutual responsibility, system-level thinking, adaptive frameworks, resilience, and shared agency between humans and AI. |
| **Autonomous Systems** | AI as a system that operates independently. Humans are bystanders or obstacles; AI is a self-willed actor. | Masks human design, deployment, oversight, and omission; erases human agency and accountability. | **Delegated authority systems** | Foregrounds human agency, choice, responsibility, and potential revocability; re-inserts the human role; frames AI as socio-technical. |
| **Intelligent Agents** | Suggests human-like reasoning, understanding, or ethical discernment. | Inflates expectations; confuses roles; muddles accountability; invites potentially irrelevant questions about personhood/agency. | **Statistical Decision Tool** | Demystifies function (pattern identification); recentres responsibility on humans; frames regulation around data, validity, and bias mitigation. |

# 9. Limitations

While this paper offers a conceptual analysis of metaphor and naming in AI and cybersecurity governance, its findings are necessarily bounded by several limitations. The analysis is largely qualitative and interpretive, drawing from examples and metaphors in policy discourse, professional practice, and media representation. As such, it does not include empirical validation through interviews, surveys, or analysis, methods that could help quantify the real-world impact of metaphorical framings on governance outcomes, institutional decisions, or public understanding.

Moreover, the linguistic critique presented here is primarily focused on English-language terminology within Global North contexts. This introduces a cultural and geopolitical bias, as naming practices and metaphorical framings may differ significantly across languages, regions, and governance systems. Without cross-linguistic or cross-cultural data, the paper's claims may not generalise to all socio-technical contexts, especially in the Global South or in low-resource language communities where AI deployment patterns and power dynamics are markedly different.

Another limitation is that the argument remains largely normative: advocating a language-first approach to governance rather than testing or implementing one in practice. Future work would benefit from piloting participatory vocabulary workshops, linguistic audits in regulatory design, or experimental interventions in governance forums to assess the feasibility and impact of these proposals, which this author strongly encourages and looks forward to hearing the outcome of these endeavours.

# 10.  Further Research

This study opens several promising avenues for future research that could expand both the empirical and theoretical dimensions of the argument. Empirically, further work could explore how specific metaphors influence behaviour and policy. Studies might employ critical discourse analysis (CDA) (Fairclough, 1992; van Dijk, 1997) or metaphor analysis (Lakoff & Johnson, 2003) to assess how terms like 'black box,' 'alignment,' or 'trustworthy AI' are deployed across regulatory texts, technical documentation, and media narratives. Interviews or ethnographies with AI developers, policymakers, and users could provide insight into how these metaphors shape expectations and accountability structures.

Cross-cultural studies are also necessary. As scholars have shown, technological governance is deeply influenced by cultural context (Jasanoff, 2005; Irani et al., 2010). Exploring how AI terminology translates (or fails to translate) across linguistic and geopolitical boundaries could illuminate hidden assumptions and structural inequalities embedded in dominant framings.

Theoretically, this work invites deeper engagement with philosophical and sociological traditions. A Foucauldian analysis, for example, could frame AI and cybersecurity metaphors as technologies of power, part of broader apparatuses of governmentality through which conduct is regulated and subjectivities are produced (Foucault, 1991). Terms like 'alignment' or 'zero trust' may function not only descriptively but prescriptively, shaping who is seen as governable or at risk.

Another approach could adopt a Latourian lens, drawing on Actor-Network Theory (ANT), to examine how metaphors function as non-human actors within socio-technical assemblages (Latour, 2005). Rather than treating naming practices as passive reflections of meaning, this perspective frames them as active nodes in a network of influence which shapes, stabilises, and translates relations between people, institutions, technologies, and ideas. In this view, metaphors help make systems durable not merely by describing them, but by enrolling allies, directing attention, and legitimising particular forms of governance.

A structuralist perspective could interpret recurring binary oppositions, such as secure/insecure, aligned/unaligned, artificial/human, as myths that societies use to resolve conceptual contradictions. These myths provide narrative closure while eliding underlying complexities (Lévi-Strauss, 1963). A structural analysis, in the Lévi-Straussian sense, might show how metaphor performs ideological work by simplifying ambiguity and concealing political stakes.

Future work must continue to examine how words encode assumptions, authorise norms, and delimit what kinds of futures can be imagined and regulated.

## Conclusion: Naming is Framing

The words we choose shape the systems we build. Misleading terms like 'cybersecurity' and 'artificial intelligence' do not just misinform the public; they misguide governance. Throughout this paper, I have offered a series of recommendations reframing techno-centric metaphors, being explicit about human roles, and co-creating vocabulary to show how linguistic clarity can lead to more inclusive, accountable, and resilient AI governance. But my intention is not just to offer a more palatable alternative, or something one could use interchangeably with the latest governance fashion. Language is not cosmetic; it is constitutional, it shapes what we govern and how.

Naming is framing, and frames determine what is visible, what is actionable and what is valued. As the Austrian-British philosopher Ludwig Wittgenstein (1922) observed, 'the limits of my language mean the limits of my world.' Therefore, if our governance vocabulary is narrow, opaque, or inherited uncritically, then the horizons of our policy imagination will be similarly constrained. To build human-centred, future-facing, and adaptive AI governance, we must patch not only the code but also the vocabulary. Naming things better is not a distraction from the 'real' work of governance; it is the real work. These linguistic patterns—whether embedded in metaphors, technical shorthand, or cultural idioms shape not only how systems are built, but also how they are governed. Before we can meaningfully regulate AI, we must regulate how we talk about it. After all, the future of AI is not artificial, and certainly not autonomous - it is us; and our first act of governance must be to govern the language we use to describe it.

# References

Agrawal, V. (2025, March 27). What is Scaffolding? *Greater Wrong.* https://ea.greaterwrong.com/posts/zfsNhHYb3P29ycjWb/what-is-scaffolding

Almada, M. (2023). Governing the Black Box of Artificial Intelligence. *SSRN Electronic Journal.* https://doi.org/10.2139/ssrn.4587609

Attard- Frost, B. (2025). Transfeminist AI Governance. *arXiv.* https://doi.org/10.48550/arXiv.2503.15682

Barrow, N. (2024). Anthropomorphism and AI Hype. *AI and Ethics,* 4, 707-711.https://doi.org/10.1007/s43681-024-00454-1

Blodgett, S., L., Barocas, S., Daume, H., & Wallach, H. (2020). Language (Technology) is Power: A Critical Survey of "Bias" in NLP. *arXiv.* https://doi.org/10.48550/arXiv.2005.14050

Brugman, B. C., Burgers, C., & Vis, B. (2019). Metaphorical Framing in Political Discourse Through Words ss. Concepts: A Meta-Analysis. *Language and Cognition*, 11(1), 41–65. https://doi.org/10.1017/langcog.2019.5

Creese, S., Dutton, W. H., & Esteve-González, P. (2021). The Social and Cultural Shaping of Cybersecurity Capacity Building: A Comparative Study of Nations and Regions. *Personal and Ubiquitous Computing*, 25(5), 941–955. https://doi.org/10.1007/s00779-021-01569-6

Cugurullo, F. (2024). The Obscure Politics of Artificial Intelligence: A Marxian Socio-Technical Critique of the AI Alignment Problem Thesis. *AI and Ethics.* https://10.1007/s43681-024-00476-9

Delgado, F., Madaio, M., Yang, S., & Yang, Q. (2023). The Participatory Turn in AI Design: Theoretical Foundations and the Current State of Practice. *arXiv.* https://doi.org/10.1145/3617694.3623261

Deroy, O. (2023). The Ethics of Terminology: Can We Use Human Terms to Describe AI? *Topoi*, 42, 881-889. https://doi.org/10.1007/s11245-023-09934-1

Duran, J., M. & Jongsma, K., R. (2021). Who is Afraid of Black Box Algorithms? On the Epistemological and Ethical Basis of Trust in Medical AI. *Med Ethics,* 47, 329-335. http://dx.doi.org/10.1136/medethics-2021-107352

Edwards, B., Hofmeyr, S., & Forrest, S. (2016). Hype and Heavy Tails: A Closer Look at Data Breaches. *Journal of Cybersecurity (Oxford)*, *2*(1), 3–14. https://doi.org/10.1093/cybsec/tyw003

Fairclough, N. (1992). Discourse and Social Change. Polity Press.

Fang, X., Li, J., Mulchandani, V., & Kim, J. (2025, February 11). *Trustworthy AI on Safety, Bias, and Privacy: A Survey.* arXiv.org. https://arxiv.org/abs/2502.10450

Foucault, M. (1991). *Governmentality*. In G. Burchell, C. Gordon & P. Miller (Eds.), The Foucault Effect: Studies in Governmentality (pp. 87–104). University of Chicago Press.

Harfouche, A., Quinio, B., & Bugiotti, F. (2023). Human-Centric AI to Mitigate AI Biases: The Advent of Augmented Intelligence. *Journal of Global Information Management*, *31*(5), 1–23. https://doi.org/10.4018/JGIM.331755

House of Commons Committee of Public Accounts. (2018). *Cyber-attack on the NHS*. UK Parliament. https://publications.parliament.uk/pa/cm201719/cmselect/cmpubacc/787/787.pdf

Irani, L., Vertesi, J., Dourish, P., Philip, K., & Grinter, R. E. (2010). Postcolonial computing: a lens on design and development. *Proceedings of the SIGCHI Conference on Human Factors in Computing Systems*, 1311–1320. https://doi.org/10.1145/1753326.1753522

Gorrieri, L. (2024). Is ChatGPT Full of Bullshit? *Journal of Ethics and Emerging Technologies 34*(1), 1-16. https://doi.org/10.55613/jeet.v34i1.149

Jasanoff, S. (Ed.). (2005). *Designs on Nature: Science and Democracy in Europe and the United States*. Princeton University Press.

Jeong, J. J., Oliver, G., Kang, E., Creese, S., & Thomas, P. (2021). The Current State of Research on People, Culture and Cybersecurity. *Personal and Ubiquitous Computing*, *25*(5), 809–812. https://doi.org/10.1007/s00779-021-01591-8

Johnson, D. G., & Verdicchio, M. (2017). Reframing AI Discourse. *Minds and Machines, 2*7(4), 575–590. https://doi.org/10.1007/s11023-017-9417-6

Kinnula, M., Livari, N., Kuure, L., & Molin-Juustila, T. (2023). Educational Participatory Design in the Crossroads of Histories and Practices – Aiming for Digital Transformation in Language Pedagogy. *Computer Supported Cooperative Work (CSCW), 32,* 745-780. https://doi.org/10.1007/s10606-023-09473-8

Lakoff, G., & Johnson, M. (2003). *Metaphors We Live By* (2nd ed.). University of Chicago Press.

Latour, B. (2005). *Reassembling the Social: An Introduction to Actor-Network-Theory*. Oxford University Press.

Lévi-Strauss, C. (1963). *Structural Anthropology*. Basic Books.

Maas, M. (2023). Concepts in Advanced AI Governance: A Literature Review of Key Terms and Definitions. *Institute for Law & AI*. http://dx.doi.org/10.2139/ssrn.4612473

Maleki, N., Padmanabhan, B., & Dutta, K. (2024). AI Hallucinations: A Misnomer Worth Clarifying. *arXiv.* https://doi.org/10.48550/arXiv.2401.06796

Makridis, C. A. (2021). Do Data Breaches Damage Reputation? Evidence from 45 Companies Between 2002 and 2018. *Journal of Cybersecurity (Oxford)*, *7*(1). https://doi.org/10.1093/cybsec/tyab021

Mizrak, F., Demirel, H. G., Yaşar, O., & Karakaya, T. (2025). Digital Detox: Exploring the Impact of Cybersecurity Fatigue on Employee Productivity and Mental Health. *Discover Mental Health*, *5*(1), 25–21. https://doi.org/10.1007/s44192-025-00149-x

National Audit Office. (2018). *Investigation: WannaCry Cyber Attack and the NHS.* https://www.nao.org.uk/wp-content/uploads/2017/10/Investigation-WannaCry-cyber-attack-and-the-NHS.pdf

NIST. (2023). *Artificial Intelligence Risk Management Framework (AI RMF 1.0)*. National Institute of Standards and Technology. https://nvlpubs.nist.gov/nistpubs/ai/NIST.AI.100-1.pdf

Papagiannidis, E., Mikalef, P., & Conboy, K. (2025). Responsible Artificial Intelligence Governance: A Review and Research Framework. *The Journal of Strategic Information Systems, 34*(2), 1-18. https://doi.org/10.1016/j.jsis.2024.101885

Rodríguez-Bermejo, D. S., Vidal, J. M., & Tapiador, J. E. (2021). The Stress as Adversarial Factor for Cyber Decision Making. *Proceedings of the 17th International Conference on Availability, Reliability and Security*, 1–10. https://doi.org/10.1145/3465481.3470047

Sasse, A., Brostoff, S., & Weirich, D. (2001). Transforming the 'Weakest Link' — a Human/Computer Interaction Approach to Usable and Effective Security. *BT Technology Journal*, 19(3). https://doi.org/10.1023/A:1011902718709

Singh, T., Johnston, A. C., D'Arcy, J., & Harms, P. D. (2023). Stress in the Cybersecurity Profession: A Systematic Review of Related Literature and Opportunities for Future Research. *Organizational Cybersecurity Journal: Practice, Process and People*, *3*(2), 100–126. https://doi.org/10.1108/OCJ-06-2022-0012

Shakespeare, W. (1899). *Romeo and Juliet*. United States: George Bell & Sons.

Stevenson, A. (2015). Cybersecurity. In Stevenson, A. (Ed.), *Oxford Dictionary of English*. Oxford University Press. https://www.oxfordreference.com/view/10.1093/acref/9780199571123.001.0001/m_en_gb099444 36

Ta, R., & Lee, N., T. (2023, October 24). How Language Gaps Constrain Generative AI Development. *The Brookings Institution*. https://www.brookings.edu/articles/how-language-gaps-constrain-generative-ai-development/

Taeihagh, A. (2025). Governance of Generative AI. *Policy & Society*. https://doi.org/10.1093/polsoc/puaf001

Teo, P. (2021, February 4). How to Make Conversational AI Work for Low-Resource Languages. *Key Reply*. https://www.keyreply.com/blog/conversational-ai-low-resource-languages#:~:text=Low%2Dreso urce%20languages%20are%20those,Western%20languages%20are%20high%2Dresource.

van Dijk, T. A. (1997). *Discourse as Structure and Process*. Sage Publications.

von Eschenbach, W. J. (2021). Transparency and the Black Box Problem: Why We Do Not Trust AI. *Philosophy & Technology*, *34*(4), 1607–1622. https://doi.org/10.1007/s13347-021-00477-0

Whelehan, A. (2025, April 11). *Karen*. Baby Name Meaning, Origin and Popularity. https://www.thebump.com/b/karen-baby-name

Wittgenstein, L. (1922). *Tractatus Logico-Philosophicus*. London: Routledge & Kegan Paul.